\documentclass{article}
\usepackage[utf8]{inputenc}
\usepackage{xcolor}
\usepackage{authblk}
\usepackage{amssymb}
\usepackage{amsmath}

\title{The Demographics of Exoplanets} 
\author[1]{B.\ Scott Gaudi}
\author[2]{Jessie L.\ Christiansen}
\author[3]{Michael R. Meyer}
\affil[1]{Department of Astronomy, The Ohio State University}
\affil[2]{Caltech/IPAC-NASA Exoplanet Science Institute, Pasadena CA}
\affil[3]{Department of Astronomy, University of Michigan}
\date{}
\usepackage{natbib}
\usepackage{graphicx}
\usepackage{geometry}
\usepackage[utf8]{inputenc}

\begin{document}
\maketitle

\section{Introduction}
The field of exoplanets can be thought of as resting on three broad pillars: demographics, characterization, and the search for habitable worlds and life\footnote{Although the study of disks is obviously strongly linked to the study of exoplanets, particularly because the properties of protoplanetary disks provide the {\it ab initio} conditions for planet formation, we will restrict our discussion in this short essay to exoplanets in systems where the planets are fully formed and the protoplanetary disk is essentially gone (e.g., ages of $\gtrsim$ 10 Myr, c.f. \cite{Yao:2018})}.  Of course, these divisions are somewhat arbitrary, the boundaries between them are generally fuzzy, and all three are interrelated.  Nevertheless, subdividing the field into these three categories of inquiry provides a useful way to frame the science goals and questions of the exoplanet field. 
In the broadest sense, the overall goal of exoplanet demographic surveys is to determine the frequency and distribution of planets as a function of as many of the physical parameters that may influence planet formation and evolution as possible, over as broad of a range of these parameters as possible.  These parameters can include, but are not limited to: 1) properties of the planets, including the planet mass $m_p$, radius $r_p$, and orbital properties (e.g., semimajor axis $a$, period $P$, eccentricity $e$); 2) properties of the host stars, including mass $M_*$, radius $R_*$, effective temperature $T_{\rm eff}$, luminosity $L_*$, metallicity [Fe/H] and detailed elemental abundances, activity, and age; and 3) environmental properties of the system, including multiplicity and birth environment.  

These distribution functions represent the ultimate empirical data set (e.g., the ``ground truth") that all theories of planet formation and evolution must reproduce.  To the extent that most planets form in a bottom-up scenario, {\it ab initio} theories of planet formation must explain how the roughly micron-sized dust grains grow by $\sim$12--14 orders of magnitude in radius and $\sim$38--40 orders of magnitude in mass to the exoplanets we detect today.  The physical mechanisms by which protoplanets accumulate mass and migrate as they cross through these many orders of magnitude in mass and radius vary dramatically, and may depend on the properties and environment of the host star, either directly or indirectly (see \cite{Raymond:2020} for a review). The net result is that signatures of these physical mechanisms should be imprinted on these planet distribution functions\footnote{The extent to which this is true may depend on the number of independent variables that contribute to the outcomes (e.g. the central limit theorem resulting in log-normal distributions), and which are scale free (e.g. resulting in power-law distributions).}.  Therefore, by comparing these planet distributions to the predictions of planet formation theories, we can begin to both test and refine these theories. 
An understanding of exoplanet demographics and the physical mechanisms that sculpted the planet distribution function is also essential for understanding planetary habitability.  Of course, the first step is to determine the frequency of potentially habitable planets, often referred to as $\eta_\oplus$\footnote{While there is no universally agreed-upon definition of $\eta_\oplus$, it generally refers to the frequency of terrestrial (rocky) planets in the traditional habitable zone of sunlike (FGK) stars.  See, e.g., \cite{Kopparapu:2013,Kopparapu:2014}.}. We discuss the most recent estimates of $\eta_\oplus$ and the opportunities for refining this estimate below. However, the habitability of any given planet can only be understood in the context of the entire planetary system, including its evolution.  This is (in part) because classical theories predicted that planets that formed in the traditional habitable zone interior to the snow line of the disk were initially dry, thereby requiring giant planets to serve as a delivery mechanism for volatile species (e.g. carbon, nitrogen, and oxygen in the form of water) from beyond the snow line \cite{Morbidelli:2016}.  On the other hand, more recent theories predict that water rich ``pebbles'' are likely to migrate interior to the snow line, which may result in the formation of many water worlds in planet population synthesis models (c.f. \cite{Lichtenberg:2019}).  Thus giant planets formed near the snow line could also act barriers to limit the inward migration of such volatile-rich material.

\section{State of the art}  
\begin{figure}
\includegraphics[width=0.515\textwidth]{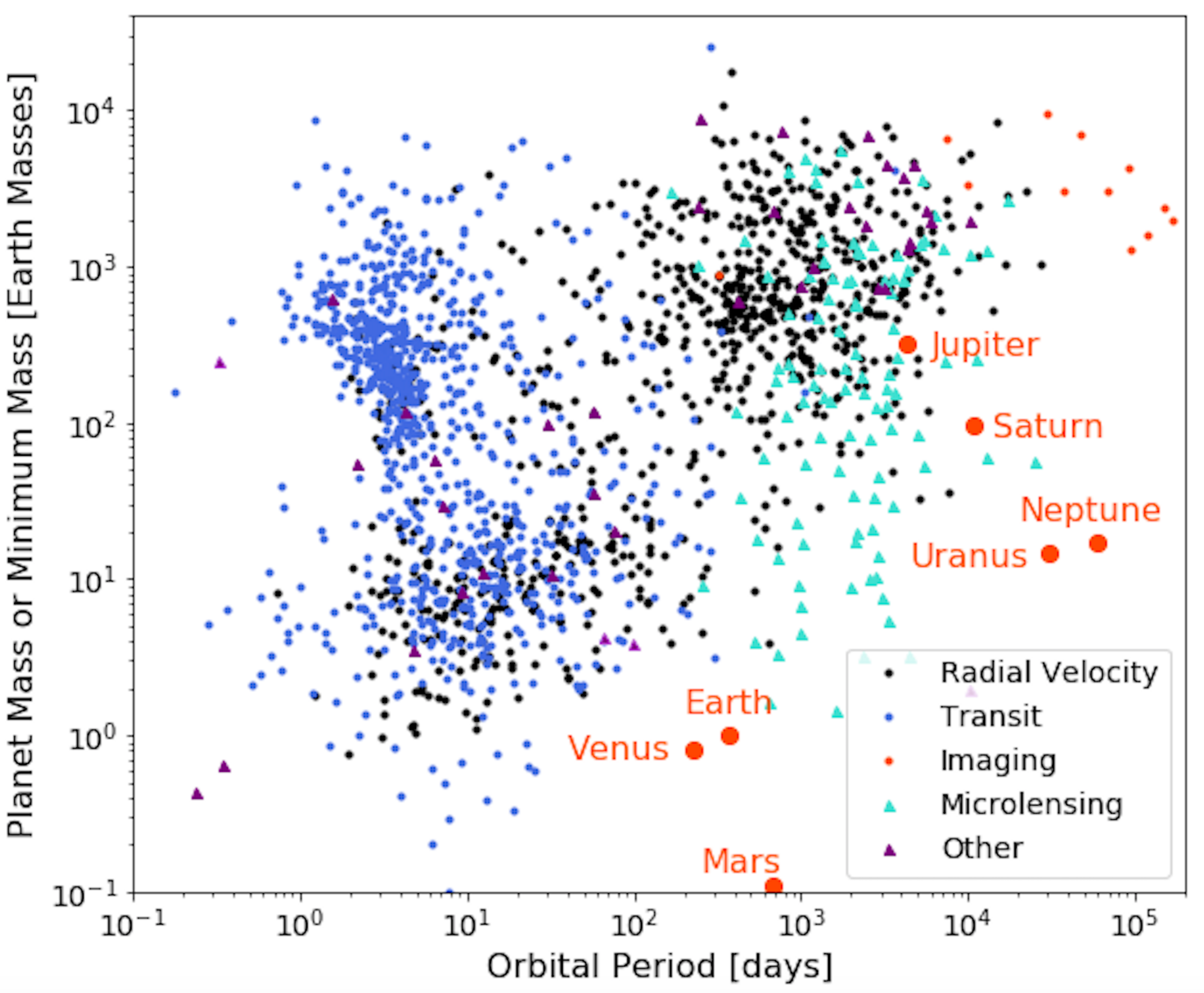}
\includegraphics[width=0.468\textwidth]{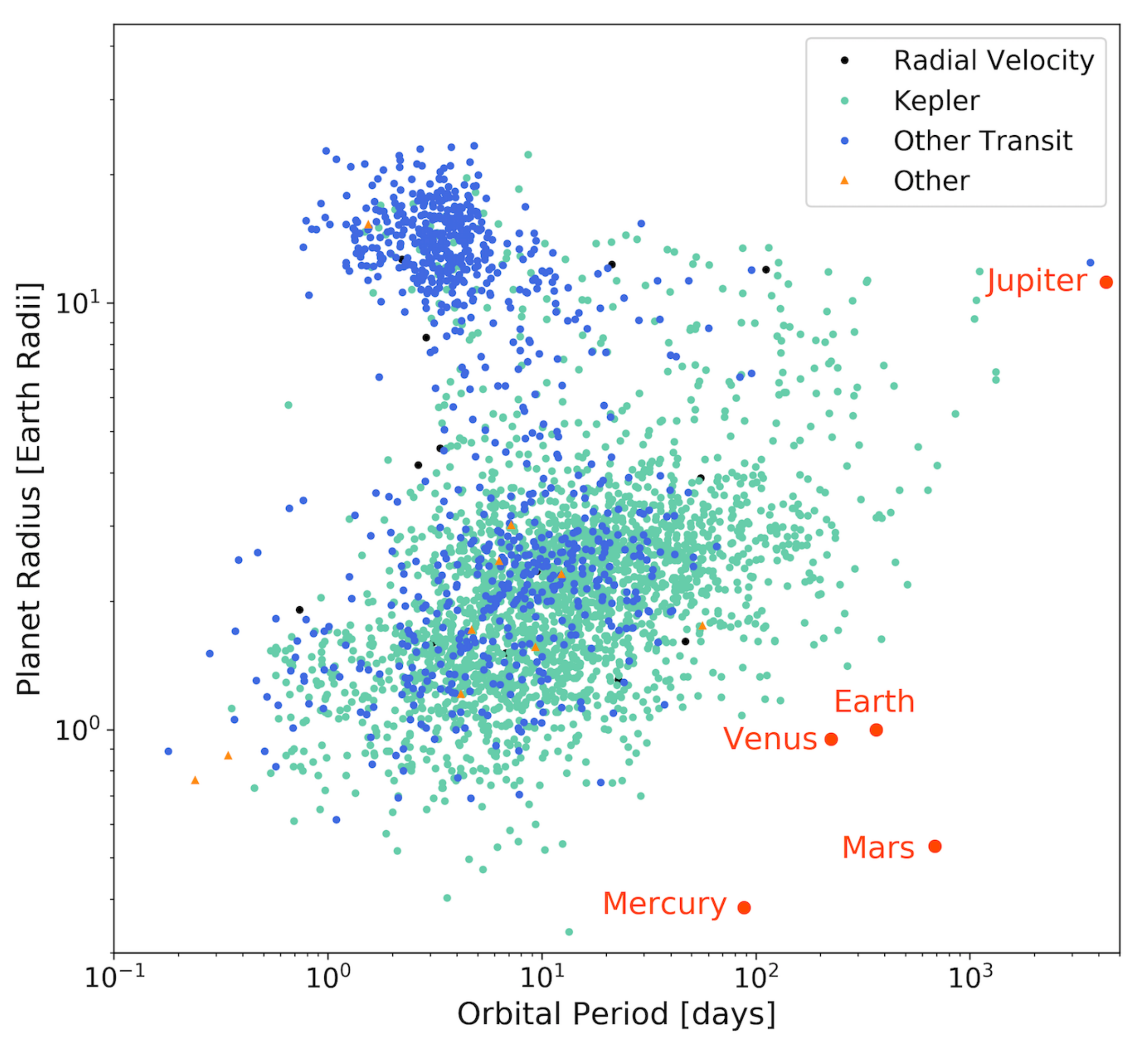}
\caption{The distributions of the $\sim$4300 confirmed low mass companions to stars as a function of their mass and period (left) and radius and period (right) where semi-major axis is converted to period Kepler's third law. The color coding denotes the method by which planets were detected.  Some of the features in these diagrams are real, however many are due to the selection effects. For example, the large population of 1--4~$R_\oplus$ planets in the right panel is largely missing in the left panel due to the fact that these planets were primarily detected by Kepler, and thus typically have host stars that are too faint to enable a measurement of their mass via radial velocity.  Similarly, the fact that there are nearly equal numbers of Hot Jupiters as cold Jupiters in the left panel is due to the fact that ground-based transit surveys, which have larger sample sizes than radial velocity surveys, are primarily sensitive to Hot Jupiters. Thus Hot Jupiters are over-represented in the figures. Finally, the paucity of planets in the lower-right corner of each plot is purely a selection effect due to the fact that the radial velocity, transit, and direct imaging methods are not currently sensitive to planets in this region of parameter space.  Note that $\sim 25$ directly-imaged planets are not shown in the left plot because they have periods that are greater than $\gtrsim 10^6$~days.   This figure is based on data from the NASA Exoplanet Archive: https://exoplanetarchive.ipac.caltech.edu/.}
\label{fig:1}
\end{figure}

Although it was not the first exoplanet to be detected \cite{Campbell:1988,Latham:1989,Wolszczan:1992}, the discovery of the Jovian companion to 51 Pegasi by \cite{Mayor:1995} ignited the field of exoplanets.  There are now over $4300$ confirmed exoplanets\footnote{As of 11/23/2020; https://exoplanetarchive.ipac.caltech.edu/}, which were primarily detected with four techniques: radial velocities, transits, microlensing, and direct imaging.  The distribution of the properties of known exoplanets in mass/period and radius/period space is shown in Figure \ref{fig:1}. 

Essentially all of the initial discoveries were made by the radial velocity (RV) method.  For many years, the primary focus of the field was on discovering new exoplanets and improving the precision of RV surveys in order to expand the region of parameter space to which the method was sensitive. Eventually, the sample of known exoplanets detected in individual RV surveys was sufficiently large that a reasonably constraining statistical analysis could be performed.  The first such statistical analysis was that of the Lick planet survey by \cite{Cumming:1999}, which contained 74 stars hosting 8 confirmed planetary companions, and several additional candidates.  These results were superseded by the analysis of the Keck RV survey by \cite{Cumming:2008}, which included 585 stars hosting 48 planets.  They fit a power-law mass-period distribution for planets with minimum mass $m_p\sin{i}>0.3~M_{\rm Jup}$ and $P<2000$ days, finding
\begin{equation}
        \frac{dN}{d\ln{m_p\sin{i}}~d\ln{P}}
\propto
    (m_p\sin{i})^{-0.31\pm 0.20}
    P^{0.26\pm 0.10},
\end{equation}
with a normalization factor such that $\sim$10\% of solar-type stars host at least one planet with mass and period in the above range. Extrapolating, they estimated that $\sim$20\% of solar-type stars host a giant planet ($>0.3~M_{\rm Jup}$ within 20 au. 
In the interim, several authors made the first explorations of the dependence of giant planet frequency on host star mass and metallicity (e.g., \cite{Gaudi:2002,Fischer:2005,Jones:2006,Johnson:2007}).

The precision of RV surveys has improved to the point that they are now sensitive to planets with mass as low as a few Earth masses with relatively short ($P\lesssim 50~{\rm days}$) periods. Furthermore, the time baseline of some of the longest-running surveys is such that they are now becoming sensitive to Jovian planets with periods approaching 30 years. Some of the most recent statistical analyses of RV surveys can be found in \cite{Howard:2010,Johnson:2010,Mayor:2011,Bonfils:2013,Fernandes:2019,Fulton:2019}. The primary conclusions of these studies are broadly consistent and can be roughly summarized as: (1) For relatively short periods and solar-type hosts (setting aside hot Jupiters), the planet mass function rises dramatically towards low-mass planets, such that sub-Neptunes and super-Earths are at least an order of magnitude more common than giant planets in the same period range; (2) The frequency of giant planets orbiting solar-type stars increases with increasing period (in agreement with \cite{Cumming:2008}) up to $\sim 5$ years ($\sim 3$~au), or roughly the location of the snow line in the solar protoplanetary disk \cite{Morbidelli:2016}, and then appears to decline with increasing separation; (3) the occurrence rate for giant planets with semimajor axes of $\lesssim 2.5$~au increases with host star mass and metallicity \cite{Fischer:2005,Johnson:2007}.  

Simultaneously, surveys using other exoplanet detection methods were being developed and conducted, including the transit, microlensing, and direct imaging methods. These surveys yielded the first detections of exoplanets by transits in 2003 \cite{Udalski:2002,Konacki:2003}), by microlensing in 2004 \cite{Bond:2004}, and by direct imaging in 2008 \cite{Marois:2008}\footnote{A directly-imaged planetary mass companion was discovered around the brown dwarf primary, 2MASS 1207 \cite{Chauvin:2004}, in 2005, but this system likely formed in manner more akin to a stellar binary, highlighting the importance of mass ratio in assessing the planetary nature of low mass companions.}.  Eventually, these methods also accrued sufficiently large samples of exoplanets to perform meaningful statistical analysis, see, e.g. \cite{Gaudi:2002,Gould:2010,Cassan:2012,Gould:2006,Mochejska:2005,Burke:2006,Hartman:2009,Masciadri:2005,Metchev:2009}.

To date, microlensing surveys have detected over 100 exoplanets. The most state-of-the art statistical analysis of the frequency of bound planets detected by microlensing was performed by \cite{Suzuki:2016}, based on the Microlensing Observations in Astrophysics II survey (MOA-II). Their sample consisted of 23 planet detections. Microlensing is generally sensitive to the planet/star mass ratio $q=m_p/M_*$ rather than the planet mass\footnote{Although, we note that it is possible to estimate the host star mass and thus the planet mass using a variety of techniques (see \cite{Bennett:2007}), and for the forthcoming survey by the Nancy Grace Roman Space telescope, such measurements should be routine \cite{Bennett:2002}.}.  In addition, it is sensitive to the instantaneous projected separation of the planet in units of the angular Einstein ring radius $s$.  Microlensing is generally sensitive to planets orbiting a range of hosts, including brown dwarfs, main-sequence stars, and stellar remnants.  Although the majority of hosts are expected to be low-mass stars, a significant fraction will be due to solar-type stars, and indeed some solar-type exoplanet hosts have already been identified (e.g., \cite{Vandorou:2020}).  Finally, microlensing surveys are most sensitive to planets with a projected separation from their parent star of roughly twice the snow line. See \cite{Gaudi:2012} for a review of microlensing searches for planets. The results from the MOA-II analysis provided strong evidence for a break in the mass ratio function (the distribution of mass ratios) at a mass ratio of $q_{\rm br}=1.7\times 10^{-4}$, roughly that of Neptune and the Sun (hereafter NSMR). Parameterizing the mass ratio $q$ and projected separation distribution, $s$, as a broken power law, they find:
\begin{equation}
   \frac{dN}{d\log{q}~d\log{s}}= 
   0.61^{+0.21}_{-0.16}\left[\left(\frac{q}{q_{\rm br}}\right)^{-0.93\pm 0.13}\Theta(q-q_{\rm br}) + \left(\frac{q}{q_{\rm br}}\right)^{0.6^{+0.5}_{-0.4}}\Theta(q_{\rm br}-q)\right]s^{0.49^{+0.47}_{-0.49}},
\end{equation}
where $\Theta(x)$ is the Heaviside step function.
This implies that the mass ratio function of cold exoplanets with mass ratios above that of the NSMR is steeper than by found by \cite{Cumming:2008}.  In addition, \cite{Suzuki:2016} found that the distribution of planets beyond the snow line is consistent with a log-uniform ({{\"O}pik} \cite{Opik:1924}) distribution.  Finally, the results of \cite{Suzuki:2016} imply that NSMR planets are likely the most common planets beyond the snow line. We note that these results were subsequently refined by \cite{Udalski:2018} and \cite{Jung:2019}.

The transit detection technique allows for a much larger number of stars to be surveyed at once, albeit with strict orientation requirements, and at greater orbital separations for low-mass planets, than the radial velocity technique (although still smaller orbital separations than are probed by microlensing). Some early transit demographics studies took advantage of this to examine nearby star clusters as laboratories for measuring occurrence rates as a function of controlled variables, e.g. metallicity and stellar density in 47 Tuc, $\omega$ Centauri, and other clusters  \cite{Mochejska:2005,Burke:2006, Gilliland:2000,Weldrake2005,Weldrake2008}. However, the first large-scale statistical studies with transiting exoplanets were performed with data from the NASA Kepler mission \citep{Borucki2010}. Prior to its launch in 2009 there were $\sim$50 known transiting planets; to date an additional $\sim 2400$ planets have come from its prime mission dataset, ushering in the statistical age of exoplanets. 

\begin{figure}[t]
\includegraphics[width=\textwidth]{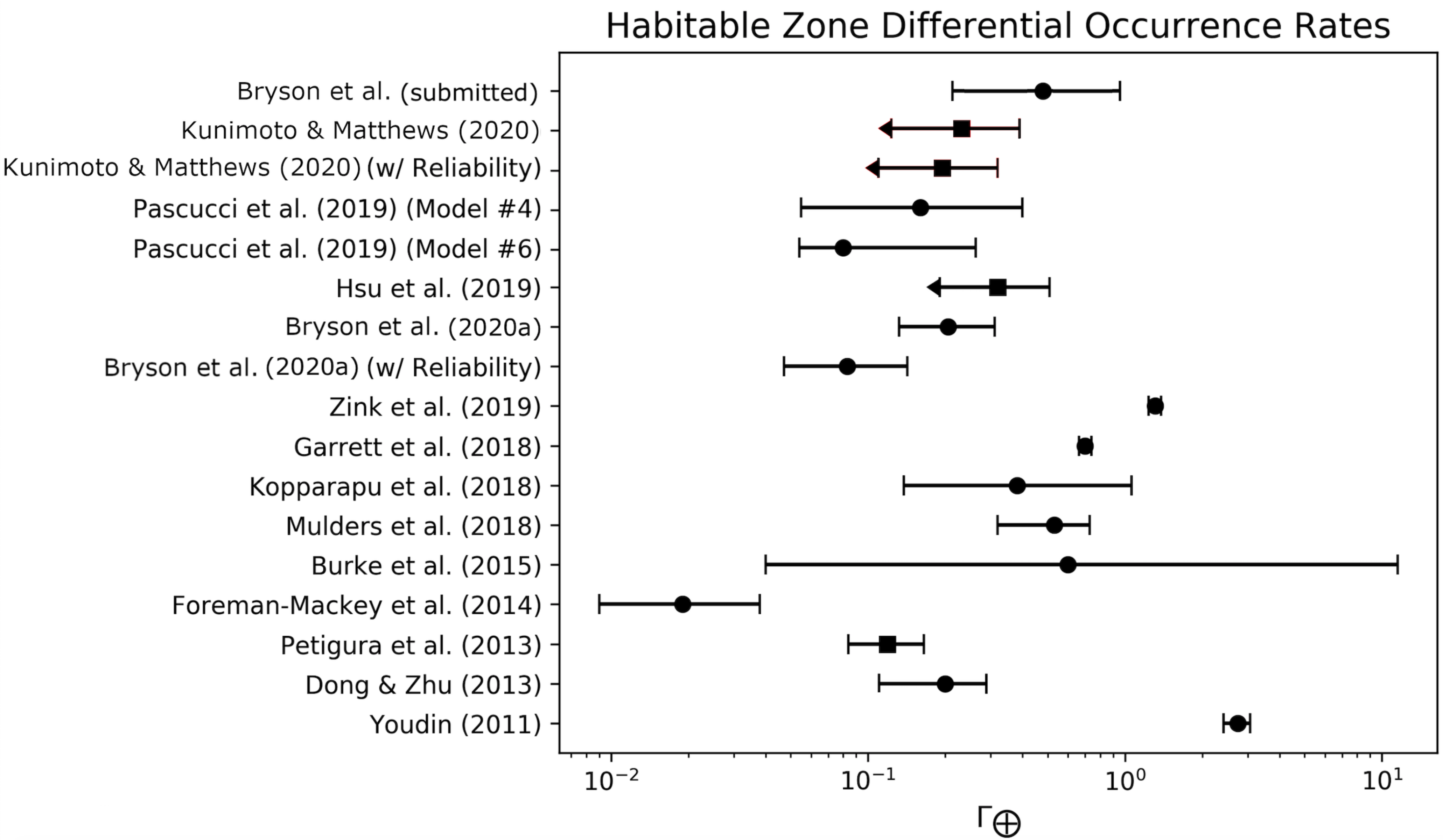}
\caption{Modified from Fig. 14 of Kunimoto \& Matthews (2020), showing a collection of $\Gamma_{\oplus}$ values from the literature: \cite{Kunimoto2020, Pascucci2019, Hsu2019, Bryson2020a, Zink:2019, Garrett2018, Kopparapu2018, Mulders2018, Burke2015, Foreman-Mackey2014, Petigura2013, Dong2013, Youdin2011}. Squares indicate that grid-based occurrence rates were explored \citep{Kunimoto2020, Hsu2019, Petigura2013}, while circles indicate a functional form was fitted and integrated for the occurrence rate (all others).}
\label{fig:2}
\end{figure}

Some important results from Kepler include the bi-modality of the size distribution of small planets \citep{Fulton2018}, the discovery of the hot Neptune desert \citep{Mazeh2016}, the clustering of planet sizes within planetary systems \citep{Weiss:2018,He2019}, and the decreasing metallicity dependence of planet occurrence rate with decreasing planet size \citep{Petigura2018}. The primary goal of the mission was to measure $\eta_{\oplus}$, however a combination of the unexpectedly high noise properties of sun-like stars and several hardware issues led to a final dataset that is incomplete in the relevant parameter space. Nevertheless, there have been many increasingly sophisticated attempts to constrain $\eta_{\oplus}$. These are summarized in Figure \ref{fig:2} via a metric labelled $\Gamma_{\oplus}$, which is the value of the occurrence rate model (whether a functional fit in radius/period space or a grid of discrete bins over radius and period) evaluated around the values for Earth ($P=365.25$ d, $R_p=1.0R_{\oplus}$). The most recent investigations from 2019 and 2020 include robust corrections for completeness and reliability, and vary mostly in the functional form they fit to the underlying population, and the parameter space over which it is fit. From a spread of over three orders of magnitude in the first few years of analysis, the community now appears to be converging on an $\eta_{\oplus}$ value between 5--50\%, with the remaining uncertainly arising largely from the small number of detections in the relevant parameter space. 

Direct imaging surveys, enabled by state-of-the-art extreme (defined by actuator density on deformable mirrors) adaptive optics (AO) on 6--12 meter telescopes, are limited to gas giants $>$1 Jupiter mass beyond 10 AU found around nearby young stars, as well as brown dwarf companions that could be a natural extension of the stellar companion mass ratio distribution (e.g., \cite{Reggiani:2016}).  Early studies made predictions of expected yields based on extrapolations of the power-law models from \cite{Cumming:2008} and fitted null results with an outer cutoff radius (\cite{Lafreniere:2014,Nielsen:2010,Heinze:2010,Vigan:2017}).   Such studies are done in the wavelength range from 1--4 microns as young planets ($<$300 Myr) are brighter and hotter than their older counterparts, and current AO systems perform well (\cite{Macintosh:2015, Beuzit:2019,Guyon:2018}).  It can be shown that the L-band (3.6-4.0 microns) is preferred for slightly older stars which enable cohorts of closer stars to comprise target samples \cite{Heinze:2010}.  
The youngest stars are rare in the solar neighborhood, therefore target stars are generally 15--150 pc away.
Results to date can be summarized as: a) gas giant planets beyond 10-50 AU are rare; b) wide orbit gas giants from 1-15 Jupiter mass appear to be more common around higher mass stars, but are also consistent with a common mass ratio distribution \cite{Vigan:2020}; c) brown dwarf companions (20-76 Jupiter masses) are more common around M dwarfs compared to higher mass stars \cite{Nielsen:2019}, though consistent with a common companion mass ratio distribution \cite{Vigan:2020}; and d) the eccentricity distribution may present a dichotomy where lower planetary mass companions have lower eccentricities compared to brown dwarf companions \cite{Bowler:2020}. 
Pioneering attempts are just being made to image planets that are in thermal equilibrium with the closest stars to the Sun (e.g. $\epsilon$ Eri b discovered by RV with imaging attempts by \cite{Mawet:2019} and references therein).  With temperatures of 100--500 K, set by stellar luminosity and orbital radius, observations are made from 5--13 microns around nearby stars with spectral type $>$K5.  However, even detecting a super-Earth around Alpha Cen A at 1.3 pc would take many nights of observtions (c.f. \cite{Kasper:2019}). 

\section{Important questions and goals} 

As is clear from Figure \ref{fig:1}, our census of exoplanets remains substantially incomplete, and as discussed below, practical obstacles have hindered attempts to combine statistical constraints from multiple surveys and detection methods.  Many major questions about the demographics of exoplanets remain.  We mention a few of the most notable here, but emphasize that this list is not meant to be complete.

\begin{itemize}
    \item {\bf How common are solar system analogs?} One of the most surprising results from Kepler is that the majority of stars appear to host relatively close-in, compact systems of super-Earths and/or sub-Neptunes (e.g., \cite{Howard:2012,Youdin2011,Dong2013,Swift:2013}). As our solar system does not host any analogues of such planets; this suggests that planetary architectures like our own (with small rocky planets in the temperate zone and gas giants beyond the ice line) may not be common.  However, given that most detection methods are not yet sensitive to analogs of the majority of the planets in our solar system, it is still not clear how rare planetary systems like ours are. Even if gas giants beyond the ice line are realized in 10 \% of systems, and the frequency of habitable zone terrestrial planets is 10 \%, these properties could be correlated such that the joint probability could be $>$ 1 \%. 
    \item {\bf How does the shape of the exoplanet mass function depend on semimajor axis and host star mass?} The exoplanet mass function (or mass ratio function) has been measured over many orders of magnitude using RV and microlensing surveys.  The radius distribution has been measured by Kepler, which can be converted to mass function by adopting a mass-radius relation (e.g., \cite{Chen:2017}).  Generally, the mass function measured by these surveys over their full range is known to be inconsistent with a single power law (e.g., \cite{Suzuki:2016,Pascucci2019}).  However, it is unclear if and how the normalization and detailed shape of the mass function (including any slopes and breaks of a piecewise power-law fit) depend on semimajor axis (particularly on either side of the snow line) and/or host star mass.
    \item {\bf Which is the more fundamental parameter that determines the final architectures of planetary systems: mass or mass ratio?} The physical mechanisms that govern planet formation and evolution can depend more directly on either the planet mass or planet/star mass ratio. For example, the critical mass for nucleated runaway gas accretion is more directly dependent on the core mass \cite{Muzino:1980,Pollack:1996,Rafikov:2006}, and thus (along with the disk lifetime) may influence the frequency of ``failed Jupiters''.  On the other hand, the final mass of a giant planet formed by core accretion likely depends on many factors, including the the Hill radius of an exoplanet, as well as the properties and lifetime of the disk \cite{Rosenthal:2020b}.  In the traditional model of solid accretion, the ``isolation mass" depends on the mass ratio (via the Hill radius).  However, when accounting for migration and/or pebble accretion, the importance of the isolation mass is less clear (see \cite{Rosenthal:2020} and references therein). Thus determining whether mass or mass ratio is the primary physical parameter that governs planet demographics in various regimes may help elucidate the physical of planet formation at various stages.  We note that there is preliminary evidence that the mass ratio function may be more universal \cite{Pascucci2019}. 
    \item {\bf How do the detailed architectures of planetary systems depend on the properties and environment of the host star?} There is substantial evidence that the frequency of relatively short-period giant planets depends on the mass and metallicity (Fe/H) of the host star (e.g., \cite{Gonzalez:1997,Fischer:2005,Johnson:2010}).  However, the precise dependence of the frequency of less-massive short-period planets on these quantities is less clear and likely more complex (e.g.,  \cite{Howard:2012,Dressing:2013,Fressin:2018,Petigura2018,Zhu:2016,Lee2015}). For wide-separation giant planets detected by direct imaging, there is some evidence for a dependence of the occurrence rate on spectral type, where higher mass stars are more likely to have gas giant companions \cite{Vigan:2020}. In the most general terms, setting aside zones of dynamical instability, the bulk frequency of planets does not appear to depend strongly on stellar multiplicity, although significant work is needed to determine the effects of stellar companions (mass, semimajor axis, and eccentricity) on the detailed architectures of planetary systems (see the discussion in \citep{Bonavita:2020} and references therein). The demographics of planetary systems as a function of stellar age has barely been explored (see \cite{Berger:2020} and references therein), primarily because of the notorious difficultly of determining precise ages of isolated field stars, particularly in the pre-Gaia era. Finally, there have been relatively few explorations of the effect of detailed abundances (e.g., [$\alpha$/Fe]) on exoplanet demographics, primarily because the host stars for the largest statistical sample of planets available, i.e., {\it Kepler}, are generally too faint for detailed abundances determinations. One well-posed question that has not yet been unambiguously answered is whether there is a minimum metallicity below which planet formation is inefficient or impossible \cite{Gilliland:2000,Sozzetti:2009,Faria:2016}.
    \item {\bf Are there multiple formation channels for giant exoplanets?} The two most common models for the giant planet formation\footnote{We will assume that low-mass planets, and in particular rocky planets, must primarily be formed in a bottom-up scenario.} are the bottom-up core accretion or nucleated instability model \citep{Pollack:1996}, and the top-down gravitational instability model \cite{Boss:1997}. Several theoretical studies have argued that the latter mechanism is only likely to operate in the outer parts of massive disks, and typically forms fairly massive giant planets \cite{Rafikov:2005,Kratter:2010}. A related question is whether or not the wide-separation giant planets that have been detected by direct imaging surveys are a distinct population of objects, or are simply the long-period, high-mass tail of planets indirectly detected by RV surveys \citep{Bowler:2016}, or the low--mass tail of the stellar companion mass ratio distribution (c.f. \cite{Vigan:2017,Vigan:2020}).  There is suggestive evidence based on segregation in the mass functions, metallicity, and orbital eccentricity distributions of giant planets that at least two formation mechanisms may operate \cite{Schlaufman:2018,Bowler:2020}.
    \item {\bf How many planets become unbound from their host stars, and what is the mass function of these ``free floating" planets?} The detection of the first Hot Jupiters provided evidence of large-scale migration of giant planets \cite{Lin:1996,Rasio:1996}, with several models to explain this migration invoking dynamical instabilities. These would not only form Hot Jupiters, but would also lead to the ejection of protoplanets or planets. Classical models of planet formation include competition and interaction of multiple planetary embryos (5-15 $M_\oplus$), some of which get ejected (e.g., \cite{Barclay:2017}).  Indeed, it has since been proposed that our own solar system may have experienced a much less quiescent formation history than has previously been assumed; such a formation history would also lead to the ejection of a significant number of planetary-mass bodies (see \cite{Raymond:2020} for a review). In addition, other mechanisms including star cluster dynamics have been proposed that can unbind planets from their host stars \cite{Adams:2006,Veras:2011}. Therefore, in principle, the abundance and mass function of free-floating planetary-mass objects can provide a constraint on models of the formation and evolution of planetary systems (as well as the low mass end of the initial mass function of brown dwarfs).
\end{itemize}

\section{Challenges} 

There are many challenges to determining the demographics of exoplanets over a broad region of parameter space.  These challenges have been identified by several authors, and methods to mitigate these challenges have also been identified.  Here we will simply provide a list of these challenges, and highlight a few studies that have attempted to address them.
\begin{itemize}
    \item {\bf Intrinsic biases and sensitivities of different exoplanet detection methods.} The primary exoplanet detection methods at our disposal (RV, transits, microlensing, direct imaging, and astrometry) are all endowed with their unique biases and planet detection sensitivities.  Fortunately, these methods are largely complementary.  As a result, by properly combining the results from diverse surveys, it is possible to derive constraints on nearly the entire region of parameter space. However, combining the results of different surveys using the same detection method, let alone combining the results of surveys based on different detection methods, is not trivial.  In particular, it is crucial that both the {\bf completeness} and {\bf reliability} of individual surveys be carefully quantified.  Furthermore, given that different methods are sensitive to different planetary properties, there are often auxiliary constraints or priors that must be adopted and/or marginalized over to combine results from different methods.  As a specific example, combining the results from transit surveys and radial velocity surveys requires adopting a mass-radius relation \cite{Chen:2017,Rogers:2015}.  An additional complication is that mass-radius relationships for exoplanets are not completely deterministic, and thus the dispersion in these relations must also be considered \citep{Wolfgang:2016}. Likely due to these challenges, there have been few attempts to combine demographic constraints from multiple detection methods, with some notable exceptions (\cite{Howard:2012,Clanton:2014,Clanton:2016,Meyer:2018}).
    \item {\bf Comparing constraints from microlensing surveys to those from other techniques.}. Microlensing is relatively unique amongst exoplanet detection methods in that (in general) the host star mass and distance is not constrained by the routine observables (specifically the microlensing light curve).  Generally, the only direct observables are the planet/host mass ratio and the projected separation of the planet in units of the angular Einstein ring of the host. However, for planetary microlensing events, it is typically possible to estimate the angular Einstein ring radius.  Furthermore, with high angular resolution observations with large telescopes with adaptive optics (e.g., Keck or Very Large Telescope), it is possible to measure the flux from the host star lens (e.g.,  \cite{Bhattacharya:2018}). Combining this with the measurement of the angular Einstein ring radius, it is possible to estimate the mass and distance to the lens, and the mass of the planet and projected separation of the planet in physical units.   
    With the Nancy Grace Roman Space Telescope (n{\'e}e WFIRST) Galactic Exoplanet survey \cite{Spergel:2015}, it will be possible to estimate the mass and distance to the primary lens, and thus the mass of the planet, for the majority of the detected planets \cite{Bennett:2007}. This will allow the direct comparison of microlensing demographics constraints to those from other techniques, and in particular the dependence of exoplanet demographics on host star mass.
    \item {\bf Determining the demographics of exoplanets as a function of the properties and environmental conditions of the host star.} It is observationally expensive to precisely constrain the parameters (e.g., mass, radius, abundance, age) and environment (e.g. multiplicity), of exoplanet host stars, particularly for relatively faint hosts. In some cases the multiplicity itself has an impact on the quality of the data one is able to obtain (e.g RV and direct imaging).  The largest statistical survey undertaken, i.e. Kepler, targeted a relatively narrow and relatively faint population of main-sequence FGK stars, whose parameters were initially derived using only broad and narrow band photometry \cite{Batalha:2010}. These photometrically-derived parameters turned out to be significantly biased, particularly for the cooler stars\cite{Mann2012}. Radial velocity surveys typically target an even narrower range of quiet FGK main-sequence stars that have been screened for low stellar activity and relatively slow rotation rates. Larger surveys like K2 and TESS will greatly expand the types of stars that are targeted for exoplanet searches, but the problem of accurately (and uniformly) deriving their stellar parameters remains. Several large-scale spectroscopic surveys (e.g. LAMOST \cite{Luo2015}, APOGEE \cite{Alam2015}, GALAH \cite{DeSilva2015}) are publicly releasing hundreds of thousands to millions of stellar spectra. In combination with the precise distances from the ESA Gaia mission, careful treatment of these spectra, including calibration across surveys to remove systematic offsets, and cross-matching with target lists of K2 and TESS, will enable us to constrain exoplanet demographics much more robustly as a function of stellar parameters in the near future.
    \item {\bf Determining the effect of exoplanet multiplicity on their demographics.}  All of the methods that have been used to detect exoplanets (pulsar timing, radial velocities, transits, microlensing, and direct imaging) have detected multi-planet systems, and there is little doubt that astrometry will also detect multi-planet systems as well; multi-planet systems therefore appear to be common. Indeed, it is not always widely appreciated that one of the most fundamental questions one can ask about exoplanet demographics, namely ``How common are exoplanets?", is not well-posed. This question could be rephrased as either ``What fraction of stars host (at least one) exoplanet?", or ``What is the average number of exoplanet per star?". These are two different questions with two distinct answers. For instance, there are a number of lines of evidence from Kepler that multi-planet systems may be distinct from systems with only one detected planet \cite{Lissauer2011,Johansen2012,Ballard2016,Zink:2019}. In addition, there is evidence that Kepler's multi-planet systems tend to be ``peas in a pod", e.g., that the planets in such systems tend to have similar radii and are regularly-spaced \cite{Weiss:2018}, or to be clustered in size and spacing \cite{He2019} (although c.f.\ \cite{Murchikova:2020,Zhu:2020,Weiss:2020}). Accounting for the effect of planetary system multiplicity may change $\eta_\oplus$ by an amount that is larger than the currently stated uncertainties on that value (see, e.g. \cite{Zink:2019}).
    \item {\bf Calibrating the mass-age-luminosity relationship of young exoplanets.}
    Observations of young, self-luminous, directly-detected planets can provide estimates of their luminosity, and temperature, derived from spectral energy distributions as well as spectroscopy, the combination of which provides a coarse radius estimate.  While some spectral features are known to be sensitive to surface gravity, inferring the mass of such directly-imaged planets requires an estimate of the age of the system, as such planets gradually cool as they radiate residual energy from formation \cite{Burrows:1997}.  With an age and luminosity, it is possible to use models to infer the mass of the planet, modulo an assumption of the specific entropy of the forming planet.  In principle, there is a unique temperature associated with a luminosity and age as a function of specific entropy.  However, these models are only now being calibrated with direct imaging observations of objects with dynamical mass estimates and known age (e.g. \cite{Snellen:2018}; c.f. section 3 of \cite{Bowler:2016}).  Legacy radial velocity surveys as well as Gaia provide opportunities to rectify the situation.  
    \item {\bf Direct Detection of Earth Analogs.} Mature planets that are in thermal equilibrium with their host star have two primary components to their spectrum: the reflected light and the thermal emission from reprocessed stellar radiation.  The amount of reflected light depends on the product of the radius and the albedo, modified by the (wavelength-dependent) scattering phase function.  The thermal emission spectrum of such planets depends on the temperature of the planet (which in turn depends on the scattering albedo).  Thus, by measuring both the reflected light and thermal emission of planets in equilibrium, it is possible to infer the radius, albedo, and temperature of mature directly-imaged planets.  Unfortunately, directly detecting mature planets in either reflected light or thermal emission from the ground will require future Extremely Large Telescopes (ELTs) with adaptive optics, and even then will only be possible for a relatively small sample of the most nearby targets.  Furthermore, the best targets for reflected light observations are planets orbiting low-mass stars, whereas the best targets for thermal emission are planets orbiting FGK stars.  Nevertheless, it may be possible to detect both signals for a handful of planets with radius $>$2 $R_\oplus$ orbiting inside the habitable zones of the most nearby stars (should the planets exist) \cite{Wang:2019}.  True Earth analogs orbiting in the habitable zones of FGK stars may be out of reach of ground-based facilities.  Detection of significant samples of such planets will require specially-designed space-based missions operating in the UV/optical/near-infrared such as the Habitable Exoplanet Observatory (HabEx, \cite{Gaudi:2020}), or the Large UV/Optical/InfraRed Mission (LUVOIR,\cite{LUVOIR:2019}), or space-based missions operating in the mid-infrared ($\sim 10~\mu{\rm m}$) such as the Large Interferometer for Exoplanets (LIFE, \cite{Quanz:2019}) (thermal emission).  We also note that direct detection of true Earth analogues (around Sun-like stars) via secondary eclipse is well beyond the capabilities of JWST.
    \item {\bf Measuring the masses of potentially habitable directly-imaged planets orbiting nearby solar-type stars.} There are several proposed mission concepts that have as one of their primary goals to directly detect and take the spectrum of terrestrial planets in the habitable zones of nearby sunlike stars (e.g., HabEx, LUVOIR, and LIFE).  The interpretation of the spectra of such planets obtained by these missions would be greatly simplified if the masses of these planets were known.  Unfortunately, the RV signal of an Earth-Sun analog is only $\sim$9 cm/s, which is a factor of 4--5 smaller than the current state-of-the-art for ground-based RV surveys.  As a result, substantial effort and investments have been put into improving the precision and accuracy of RV surveys to $\sim$10 cm/s and below \cite{Fischer:2016,Plavchan:2015}.  The astrometric signal of an Earth-Sun analog is only $\sim$0.3~$\mu{\rm as}$, which is almost certainly not achievable from the ground.  Several mission concepts studies that may enable astrometric precisions well below these levels have been proposed (e.g., \cite{Malbet:2016}). 
\end{itemize}

\begin{figure}[t]
\includegraphics[width=0.45\textwidth]{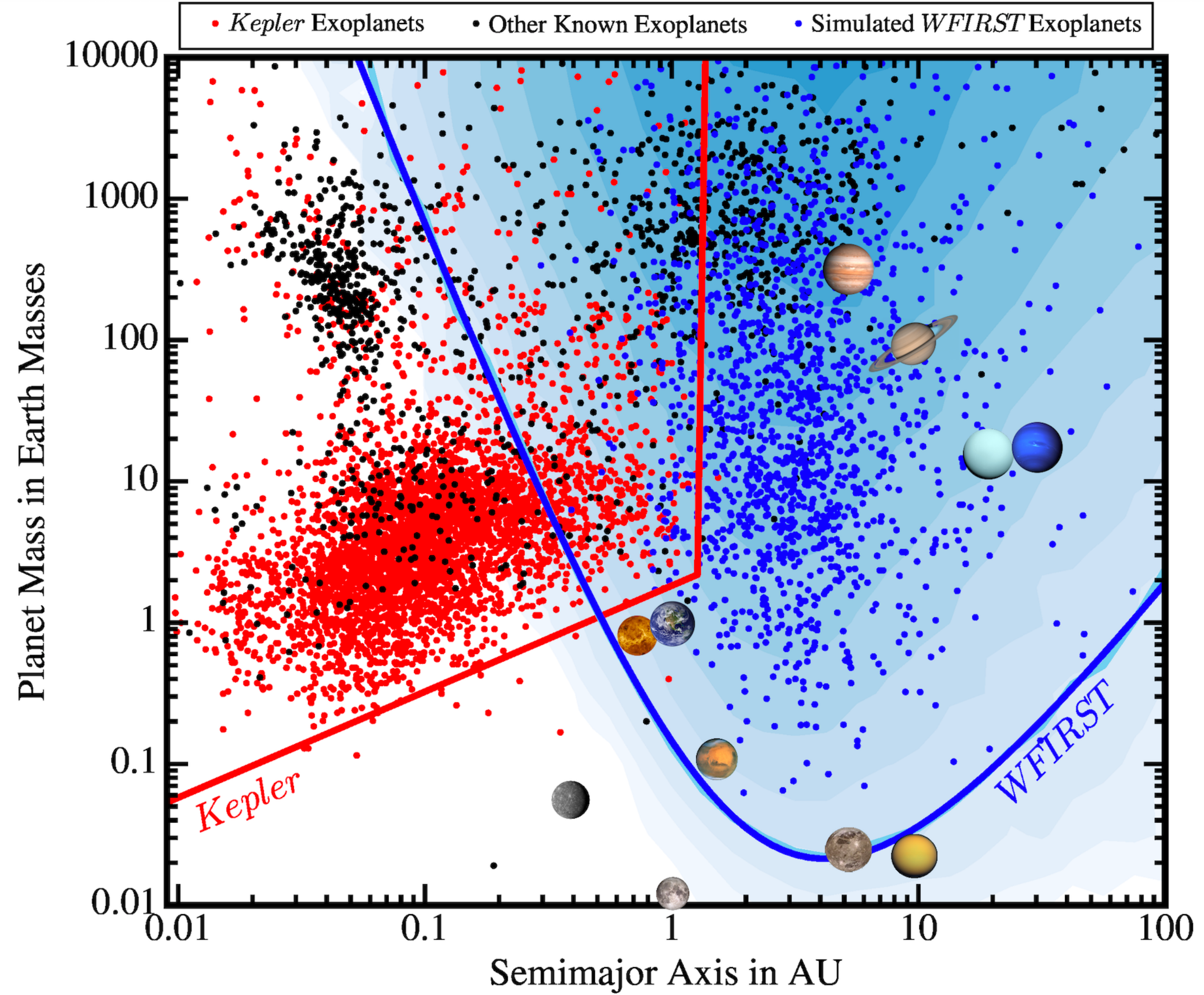}
\includegraphics[width=0.50\textwidth]{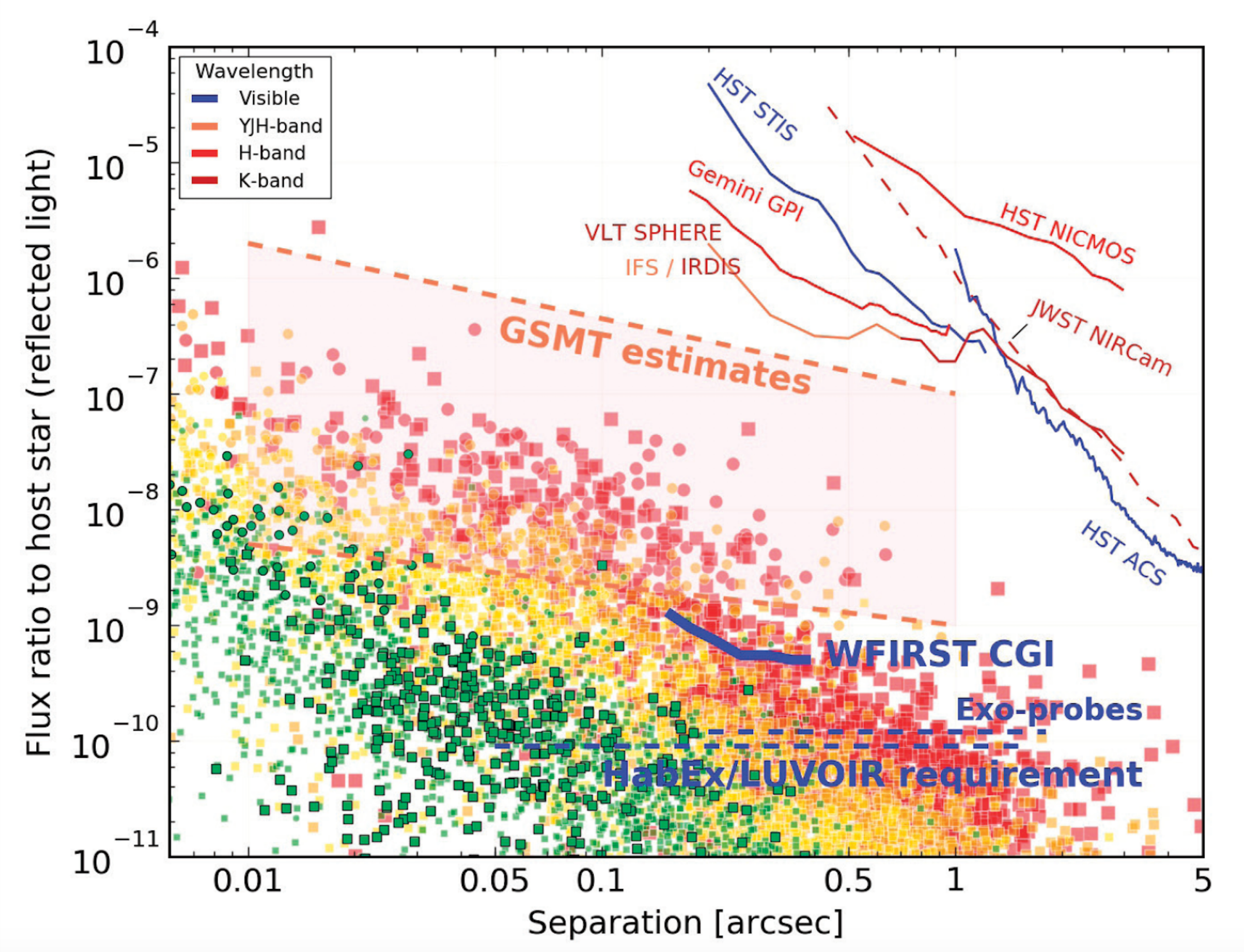}
\caption{(left) The region of sensitivity of the {\it Kepler} prime mission (red solid lines) versus the predicted region of sensitivity of the {\it Roman} (n{\' e}e WFIRST) Galactic Exoplanet Survey (blue solid lines). From \cite{Penny:2019}. Courtesy of M. Penny. © AAS. Reproduced with permission. (right) Reflected light flux ratio versus angular separation for current and future direct imaging surveys. The markers denote simulated planet populations within 27 pc using occurrence rates from \cite{Kopparapu:2018}. Various cuts have been applied to more realistically portray the detectability of these populations.
The marker size is proportional to planet
size and colors denote specific bins: giant planets (red); Neptunes (orange), mini-Neptunes (yellow); super-Earths (dark green); temperate Earths and super-Earths (light green).  Round points indicate
planets orbiting cool stars ($T_{\rm eff} < 4000$K), while square points denote planets around warmer stars. From Fig.\ 4.3 in \cite{NASESS:2018}; see that reference for additional details. Republished with permission of the National Academies Press; permission conveyed through Copyright Clearance Center, Inc.  Courtesy of D. Mawet, B. Macintosh, T.
Meshkat, V. Bailey, and D. Savransky. }
\label{fig:3}
\end{figure}

\section{Opportunities} 

In this section, we briefly summarize some of the future opportunities for refining and expanding the demographics of exoplanets.  In particular, we will focus on the prospects for exploring the region of planet parameter space (planet mass, radius, semimajor axis, and period) for which exoplanet demographics can be determined, as well as the region of host star parameter space (mass, temperature, abundance, multiplicity, and age).  As we are focusing on demographics, we will {\bf not} discuss the many exciting future prospects for detailed atmospheric characterization of exoplanets using, e.g., current and future ground-based facilities (including ELTs), current and future space-based facilities such as the Hubble Space Telescope (HST), the CHaracterising ExOPlanet Satellite (CHEOPS), the James Webb Space Telescope (JWST), Twinkle, the Nancy Grace Roman Space Telescope (Roman), and the Atmospheric Remote-sensing Infrared Exoplanet Large-survey (ARIEL), or proposed missions such as Origins.   

\begin{itemize}
    \item {\bf The Transiting Exoplanet Survey Satellite (TESS):} The NASA TESS mission \cite{Ricker2015} will observe almost the entire sky at 30-minute cadence with observing baselines of 27--351~d; transiting exoplanet detections are expected down to $T_{\rm mag}=14$.  While the range in planet radius and period covered by TESS is generally narrower than that covered by Kepler, the target list is several orders of magnitude larger. Thus, although the primary goal of TESS is to provide a sample of small ($<4R_\oplus$) transiting planets orbiting bright stars, it will nevertheless contribute significantly to our understanding of exoplanet demographics. In particular, the TESS target sample includes stars covering a much larger range of stellar properties.  This fact, combined with the fact that the host stars will be relatively bright and thus amenable to detailed characterization, will enable more in-depth investigations of planet demographics as a function of, e.g., stellar age, mass, temperature, and metallicity. This information will provide important constraints on the dominant physical processes sculpting the planet distributions we see. In addition, the large sample of planets with well-constrained masses and radii amenable to atmospheric characterization will allow us to expand our demographic studies to include atmospheric properties.

\item {\bf Gaia:} ESA's Gaia mission possesses revolutionary capabilities to discover planets down to Neptune masses around nearby low mass stars at relatively large orbital radii via astrometry.  Previous estimates that extrapolate radial velocity distributions \cite{Cumming:2008} predict Gaia will discover thousands of planets below a few Jupiter masses (e.g. \citep{Perryman:2014}).  However, if the orbital radius distributions peak at $<$ 10 AU these estimates are likely too optimistic. Nonetheless, Gaia will provide dozens to hundreds of planets with dynamical masses and full Keplerian orbital elements (up to a two-fold degeneracy).  These systems can be used to explore the architectures and (in particular) the coplanarity of planetary systems.  Furthermore, many of these systems will be amenable to characterization by direct imaging, offering unprecedented opportunities to compare the properties of wide-orbit planets with close--in planets discovered via transit.  

\item {\bf JWST:} In the background limit, JWST's ability to directly detect exoplanets will outperform ground-based telescopes by orders of magnitude in the infrared, enabling the discovery of planets $>$ 10 Earth masses around nearby (10-50 pc), young (30-300 Myr), late-type stars ($>$ M0).  However, in the contrast--limit, it is likely that ground-based telescopes with extreme AO will outperform JWST within $\sim$ 10 $\lambda$/D.  Spectral retrieval studies will benefit tremendously from the higher signal--to--noise ratio and wavelength coverage of JWST relative to ground-based telescopes, providing constraints on volatile composition to confront planet formation theory. 

\item {\bf Extremely Large Telescopes (ELTs):} ELTs will provide the angular resolution and sensitivity to detect planets below 1 Jupiter mass around nearby stars, many of which will have dynamical masses from radial velocity measurements or Gaia.  This will be the first opportunity to characterize a large sample of mature wide-orbit planets to compare with gas giants found around young stars.  ELTs will enable imaging at angular scales around stars close enough that imaging and radial velocity (and astrometric) samples will have large overlap in orbital radius, on both sides of the local maxima between 1--10 AU. In addition, ELTs will be able to directly detect planets down to $<$2 Earth radii around the very nearest stars, including possibly in the habitable zone.  Thermal infrared imaging will prefer stars earlier than K5 while reflected light studies will be best around M dwarfs, particularly as AO performance pushes into the visible (e.g., \cite{Males:2019}). 

\item {\bf The Nancy Grace Roman Space Telescope (Roman):} Roman will perform the Roman Galactic Exoplanet Survey (RGES) using the microlensing detection method.  Initial estimates from the RGES survey predict that it will detect $\sim$1500 cold planets (planets beyond the snow line), with masses down to the mass of Ganymede \cite{Penny:2019}.  When combined with {\it Kepler}, Roman will complete the statistical census of planets with masses and radii greater than roughly that of the Earth, for semimajor axes from zero to infinity; see Figure 3. Indeed, Roman will be sensitive to free-floating planets (planets that are not bound to any star) with masses down to the masses of Mars \cite{Johnson:2020}. Together, Kepler and Roman will provide the empirical ground truth that any {\it ab initio} planet formation theories must match. Roman will also be able to constrain the frequency of potentially habitable planets orbiting sunlike stars ($\eta_\oplus$), complementing and bolstering the estimates from Kepler.

\item {\bf The PLAnetary Transits and Oscillations of stars (PLATO) mission:} PLATO \cite{Rauer:2014} is a selected medium-class ESA mission whose objective is to discover and characterize a large number of transiting exoplanets, as well as perform asterosiesmology on a large number of stars, including the host stars of a significant fraction of the detected transiting planets.  PLATO can be thought of as a next-generation Kepler-like survey: in particular, it will improve upon the primary Kepler survey by observing a larger area of the sky. Therefore, PLATO's target stars will typically be brighter than that of Kepler, thereby facilitating follow-up of the planetary systems, including the ability to measure masses of detected transiting planets with radial velocities. Notably, PLATO will enable a detailed exploration of planet demographics as a function of the stellar age, accessible with asteroseismology, and may also provide an improved estimate of $\eta_\oplus$.

\item {\bf Space-based direct imaging surveys:} Future proposed direct-imaging surveys, such as HabEx \cite{Gaudi:2020}, LUVOIR \cite{LUVOIR:2019}, and LIFE \cite{Quanz:2019} will enable a direct census of nearby ($d \le 10~{\rm pc}$) planetary systems over nearly the entire range of semimajor axes spanned by the planets in our solar system (see Fig.~\ref{fig:3}b for the parameter space to which HabEx and LUVOIR are expected to be sensitive).  In particular, these missions would enable the detection, as well as orbital and atmospheric characterization, of analogs of our own Venus, Earth, and Jupiter.  Depending on the precise architecture and starlight suppression technique, these missions may also be able to detect, and potentially characterize, analogs to Mars and Saturn.  Depending on the contrast floor and outer working angle, they may also be able to detect and possibly characterize gas giants at separations corresponding to the ice giants in our solar system (see, e.g., \cite{Gaudi:2020}).  In other words, these missions have the potential to detect {\bf and} characterize analogs to nearly all of the planets in our solar system, {\it in individual systems}.  As a result, these missions will enable us to ask such fundamental questions as: ``What is the conditional probability that, given the existence of a terrestrial world in the habitable zone of a nearby star, that the system hosts a giant planet beyond the snow line?" 
\end{itemize}

\section*{Epilogue}
The future of exoplanet demographics looks bright.  Of the three pillars of the study of exoplanets outlined in the introduction of this chapter, exoplanet demographics is likely the first to reach fruition.  This is as it should be.  The other two pillars of exoplanet studies: detailed characterization of exoplanets, and the search for habitable worlds and life, fundamentally rely on our basic knowledge of the architectures of planetary systems, and how these architectures depend on the host star properties.

\section*{Acknowledgements}
We would like to thank Radek Poleski for assistance with improving Figure \ref{fig:1}.  
Partial support for B.S.G. was provided by the Thomas Jefferson Chair for Space Exploration endowment at the Ohio State University. This research has made use of the NASA Exoplanet Archive, which is operated by the California Institute of Technology, under contract with the National Aeronautics and Space Administration under the Exoplanet Exploration Program.

\bibliographystyle{plain}
\bibliography{exofrontiers_exoplanetary_demographics}
\end{document}